\documentclass[preprint]{aastex63}
\usepackage{amsmath,amssymb,CJK,soul}

\newcommand\hst{\textit{HST}}

\received{--}
\revised{--}
\accepted{--}
\submitjournal{AJ}

\shorttitle{Disintegration of C/2019 Y4 (ATLAS): I. \hst~Observations}
\shortauthors{Ye et al.}

\begin{document}
\begin{CJK*}{UTF8}{gbsn}

\title{Disintegration of Long-Period Comet C/2019 Y4 (ATLAS): I. Hubble Space Telescope Observations}

\correspondingauthor{Quanzhi Ye}
\email{qye@umd.edu}

\author[0000-0002-4838-7676]{Quanzhi Ye (叶泉志)}
\affiliation{Department of Astronomy, University of Maryland, College Park, MD 20742, USA}

\author{David Jewitt}
\affiliation{Department of Earth, Planetary and Space Sciences, UCLA, Los Angeles, CA 90095-1567, USA}
\affiliation{Department of Physics and Astronomy, UCLA, Los Angeles, CA 90095-1547, USA}

\author[0000-0001-9067-7477]{Man-To Hui (许文韬)}
\affiliation{State Key Laboratory of Lunar and Planetary Science, Macau University of Science and Technology, Macau, China}
\affiliation{Institute for Astronomy, University of Hawaii, Honolulu, HI 96822, USA}

\author[0000-0002-6702-191X]{Qicheng Zhang}
\affiliation{Division of Geological and Planetary Sciences, California Institute of Technology, Pasadena, CA 91125, USA}

\author[0000-0001-6608-1489]{Jessica Agarwal}
\affiliation{Institut f\"ur Geophysik und extraterrestrische Physik, Technische Universit\"at Braunschweig, 38106 Braunschweig, Germany}

\author[0000-0002-6702-7676]{Michael S. P. Kelley}
\affiliation{Department of Astronomy, University of Maryland, College Park, MD 20742, USA}

\author{Yoonyoung Kim}
\affiliation{Institut f\"ur Geophysik und extraterrestrische Physik, Technische Universit\"at Braunschweig, 38106 Braunschweig, Germany}

\author{Jing Li (李京)}
\affiliation{Department of Earth, Planetary and Space Sciences, UCLA, Los Angeles, CA 90095-1567, USA}

\author[0000-0002-3818-7769]{Tim Lister}
\affiliation{Las Cumbres Observatory, 6740 Cortona Drive, Suite 102, Goleta, CA 93117, USA}

\author{Max Mutchler}
\affiliation{Space Telescope Science Institute, Baltimore, MD 21218, USA}

\author{Harold A. Weaver}
\affiliation{Johns Hopkins University Applied Physics Laboratory, Laurel, MD 20723, USA}



\begin{abstract}
Near-Sun Comet C/2019 Y4 (ATLAS) is the first member of a long-period comet group observed to disintegrate well before perihelion. Here we present our investigation into this disintegration event using images obtained in a 3-day {\it Hubble Space Telescope} (\hst) campaign. We identify two fragment clusters produced by the initial disintegration event, corresponding to fragments C/2019 Y4-A and C/2019 Y4-B identified in ground-based data. These two clusters started with similar integrated brightness, but exhibit different evolutionary behavior. C/2019 Y4-A was much shorter-lived compared to C/2019 Y4-B, and showed signs of significant mass-loss and changes in size distribution throughout the 3-day campaign. The cause of the initial fragmentation is undetermined by the limited evidence but crudely compatible with either the spin-up disruption of the nucleus or runaway sublimation of sub-surface supervolatile ices, either of which would lead to the release of a large amount of gas as inferred from the significant bluing of the comet observed shortly before the disintegration. Gas can only be produced by the sublimation of volatile ices, which must have survived at least one perihelion passage at a perihelion distance of $q=0.25$~au. We speculate that Comet ATLAS is derived from the ice-rich interior of a non-uniform, kilometer-wide progenitor that split during its previous perihelion. This suggests that comets down to a few kilometers in diameter can still possess complex, non-uniform interiors that can protect ices against intense solar heating.
\end{abstract}



\section{Introduction} \label{sec:intro}

Catastrophic disintegration is a common end state for comets \citep[cf.][]{Hughes1990, Chen1994}. While the disintegration of smaller, sub-kilometer-sized comets usually results in the comet turning into a cloud of dust that effectively marks its end-of-life, disintegration of multi-kilometer-sized comets can produce multiple active fragments that remain observable as distinct comets over extended periods. These fragments have orbits resembling their progenitor, and are collectively known as a comet group or comet family.

Although comet splitting and disintegration are common, only a handful comet groups/families have been identified, with most being Jupiter-family comets \citep{Boehnhardt2004, Fernandez2009}. Only two long-period comet (LPC) families have been unambiguously identified: the well-known Kreutz sungrazing comet family which contains over 4,000 known fragments \citep{Marsden1967, Marsden1989, Sekanina2004, Knight2010, Battams2017}, as well as the Liller--Tabur--SWAN group that includes at least C/1988 A1 (Liller), C/1996 Q1 (Tabur), and C/2015 F3 (SWAN) \citep[e.g.][]{Sekanina1997, Sekanina2016}.

Comet C/2019 Y4 (ATLAS) was discovered by the Asteroid Terrestrial-impact Last Alert System  \citep[ATLAS;][]{Tonry2018ATLAS} program on 2019 December 28, and was immediately noted by M. Meyer for an orbit that closely resembles another LPC, C/1844 Y1 (Great Comet)\footnote{See M. Meyer, \url{https://groups.io/g/comets-ml/topic/69345078}; as well as Minor Planet Electronic Circular 2020-A112, \url{https://minorplanetcenter.net/mpec/K20/K20AB2.html}.}. Further investigation by \citet{Hui2020} supports the idea that the two are the products of a larger comet that likely split during its last perihelion passage $\sim$5~kyr ago. The comet pair shares a perihelion distance of $q=0.25$~au and an inclination of $i=45^\circ$.

Comet ATLAS brightened rapidly from 2020 February to March, and then faded slowly, despite still being $\sim2$~months from perihelion. Clear signs of disintegration were first noted in early April of 2020 \citep[e.g.][]{Lin2020, Venkataramani2020, Ye2020a}. Additional follow-up observations showed continued fragmentation into 2020 May as the comet was moving towards perihelion and into solar conjunction, and the fragment swarm was still visible in early June as it transited through Solar Terrestrial Relations Observatory (STEREO)'s Heliospheric Imager \citep{Knight2020}, a space-based instrument that monitors the sky near the Sun.

Members of comet groups, being products of a disintegrated parent, are also prone to disintegrate. For instance, many Kreutz comets disintegrate shortly before their extreme perihelion at 0.005~au. Before Comet ATLAS, no fragmented LPC members have ever been observed to disintegrate well ($\gtrsim1$~au) before perihelion. This is not surprising given that surviving fragments should preferentially contain the most resilient constituents that have already experienced the intense heating at perihelion at least once before. It would appear unlikely that one would disrupt without being exposed to at least a comparable level of heating. The case of Comet ATLAS is thus particularly interesting, as it is the very first fragmented LPC member to disintegrate long before perihelion. Why and how did it happen, and what sets Comet ATLAS apart from other LPCs?

\section{Observations} \label{sec:obs}

We secured three orbits of the \textit{Hubble Space Telescope} (\hst) through General Observer programs 16089 and 16111. Images of Comet ATLAS were obtained using the Wide Field Camera 3 (WFC3) on 2020 April 20 and 23 (Table~\ref{tbl:obs}). The first two orbits were scheduled on 2020 April 20, separated by about three hours; the third orbit was scheduled on 2020 April 23. In each orbit we obtained five exposures ranging from 385--397 seconds. All observations were made with the telescope tracking at the comet's telescope-centric motion rate, resulting in trailed and slightly curved background stars. Exposures were dithered in order to help minimize the impact from bad pixels and detector gaps. For maximum sensitivity, all exposures were obtained through the ultra-broad F350LP filter which has a central wavelength at $\sim580$~nm and a full-width-half-maximum (FWHM) of $\sim490$~nm. Due to programmatic issues, the comet was positioned on the UVIS1 detector in the 16089 exposures and the UVIS2 detector in the 16111 exposures, causing the 16111 exposures to cover only up to $\sim1'$ tailward of the nucleus, compared to $\sim2.5'$ in the 16089 exposures. Nearly all (except one) of the fragments are within $1'$ of the nucleus, hence the impact of this programmatic difference is minimal. 

\begin{table*}
\centering
\caption{Summary of the \hst~observations. Listed are the date, time, program IDs, and filters of the observations, heliocentric distance $r_\mathrm{H}$, geocentric distance $\varDelta$, phase angle $\alpha$ of the comet, and the exposure strategy.}
\begin{tabular}{lcccccc}
\hline
 Date and time (UTC) & Program ID & Filter & $r_\mathrm{H}$ (au) & $\varDelta$ (au) & $\alpha$ & Exposure \\
\hline
2020 April 20 10:14--10:51 & 16089 & F350LP & 1.102 & 0.978 & $57.4^\circ$ & $5\times6.4$~min \\
2020 April 20 13:31--14:07 & 16111 & F350LP & 1.099 & 0.977 & $57.5^\circ$ & $5\times6.6$~min \\
2020 April 23 09:48--10:23 & 16111 & F350LP & 1.041 & 0.964 & $60.1^\circ$ & $5\times6.6$~min \\
\hline
\end{tabular}
\label{tbl:obs}
\end{table*}

Images were cleaned against cosmic rays and hot pixels using an optimized version of the {\tt L.A. Cosmic} algorithm \citep[][see also C. McCully, \url{https://github.com/cmccully/lacosmicx}]{vanDokkum2001}. Astrometric solutions were recomputed using field stars in order to enable precise astrometry of the fragments. The update of the astrometric solution is necessary because the released data only includes crude solution computed from the guide stars, which contains errors up to a few $0.1''$, or tens of WFC3 pixels. (\hst's data pipeline can compute astrometric solutions using field stars, but this capability does not extend to images with trailed field stars.) We measure the ends of the star trails, match them to the Gaia Data Release 2 (Gaia--DR2) catalog \citep{Gaia2018} using {\tt SCAMP} \citep{Bertin2010}, and re-derive the astrometric solution. The error of the updated solution is within $0.02''$, or half of one WFC3 pixel.

\begin{figure*}[!htb]
\centering
\includegraphics[width=\textwidth]{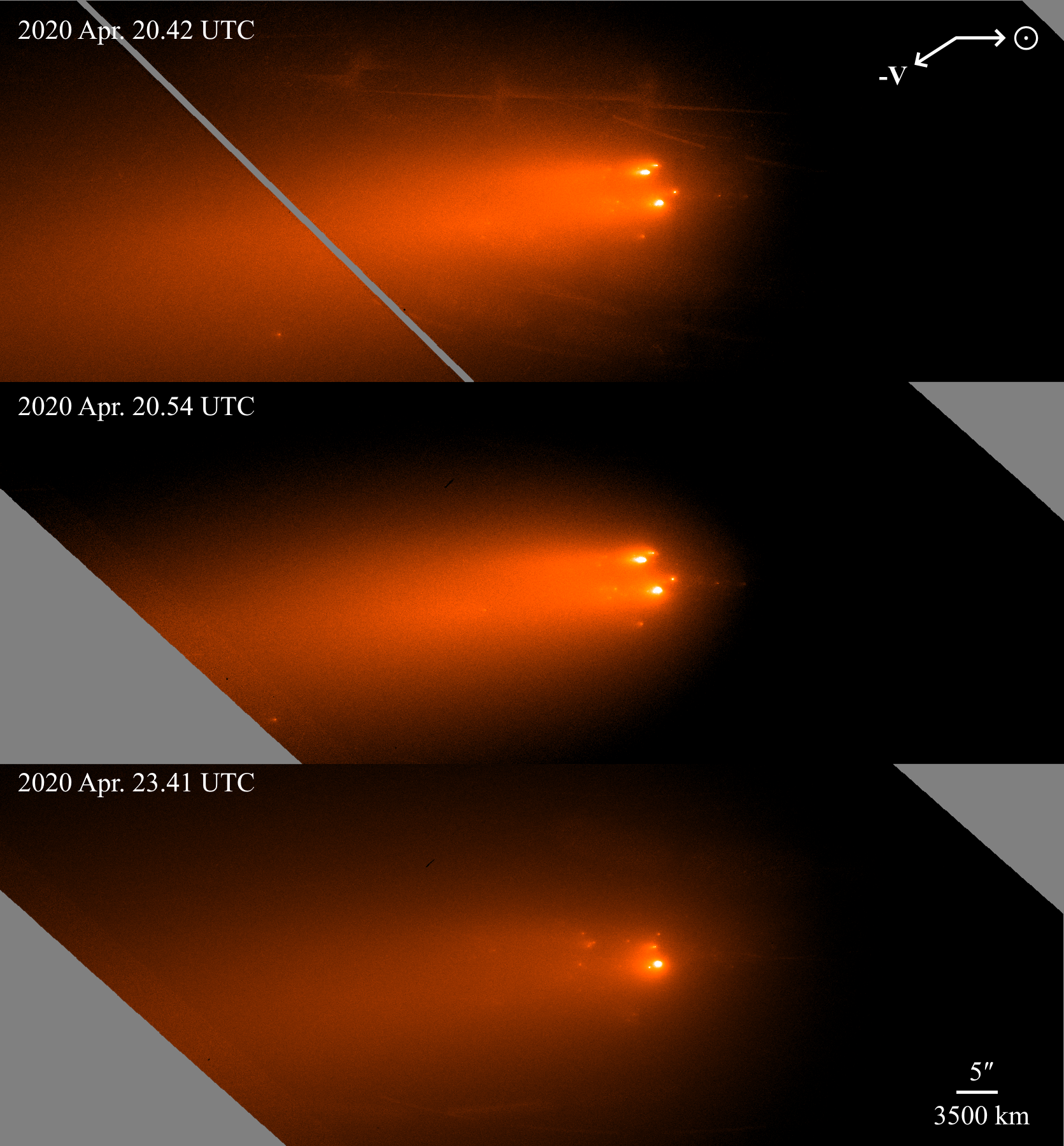}
\caption{Median-combined \hst~images from all three orbits. The arrows mark the comet--Sun vector (arrow to $\odot$) and negative velocity vector (arrow to {\bf -V}); the change of these vectors across the epochs is negligible. The diagonal streak in the upper panel is the chip gap between the two UVIS detectors. Other fainter streaks in the images are trailed background stars.}
\label{fig:mosaic}
\end{figure*}

We then median combined exposures from the same orbit into a composite image using {\tt Montage} \citep{Jacob2010}, shown in Figure~\ref{fig:mosaic}. However, we soon realized that the relative motion between fragments was so high as to be readily visible within an orbit. We estimate that relative motions between fragments were up to $0.6''$/hour, or $\sim10$~pixels in one orbit ($\sim50$~minutes of usable time). Mitigation of this effect will be discussed in the following section.

\section{Analysis}

\subsection{Fragment Identification and Measurement}
\label{sec:an:id}

\begin{figure*}[!htb]
\centering
\includegraphics[width=\textwidth]{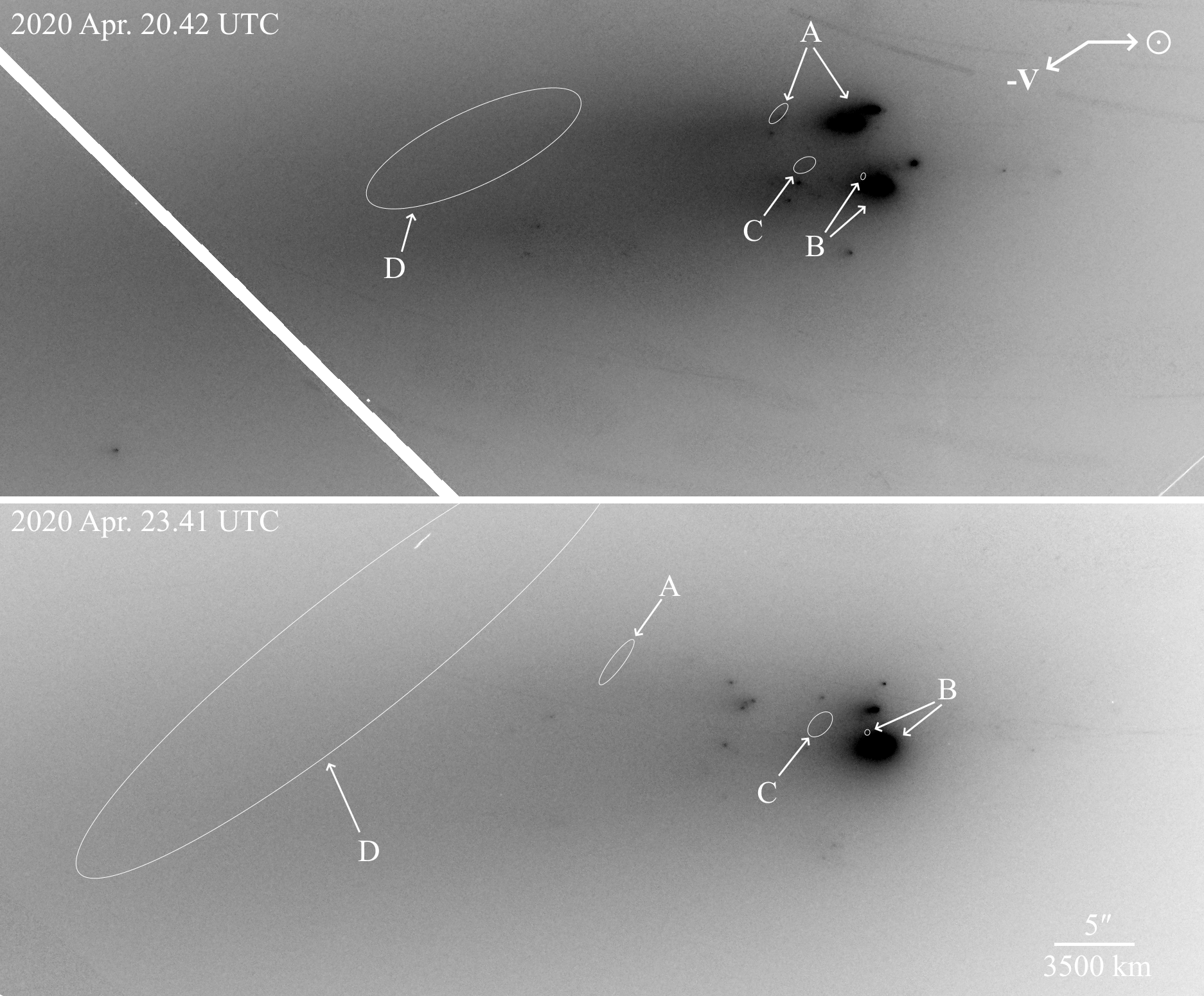}
\caption{Predicted positions and $3\sigma$ uncertainty ellipses of the four published fragments of Comet ATLAS superimposed to the composite \hst~images. The predictions are based on the ephemeris derived from JPL orbital solutions \#6 (for fragment A) and \#7 (for fragments B through D). Arrows mark the ephemeris ellipse and the corresponding fragments on actual images where applicable (see the discussion in the main text).}
\label{fig:unc}
\end{figure*}

\begin{figure*}[!htb]
\centering
\includegraphics[width=\textwidth]{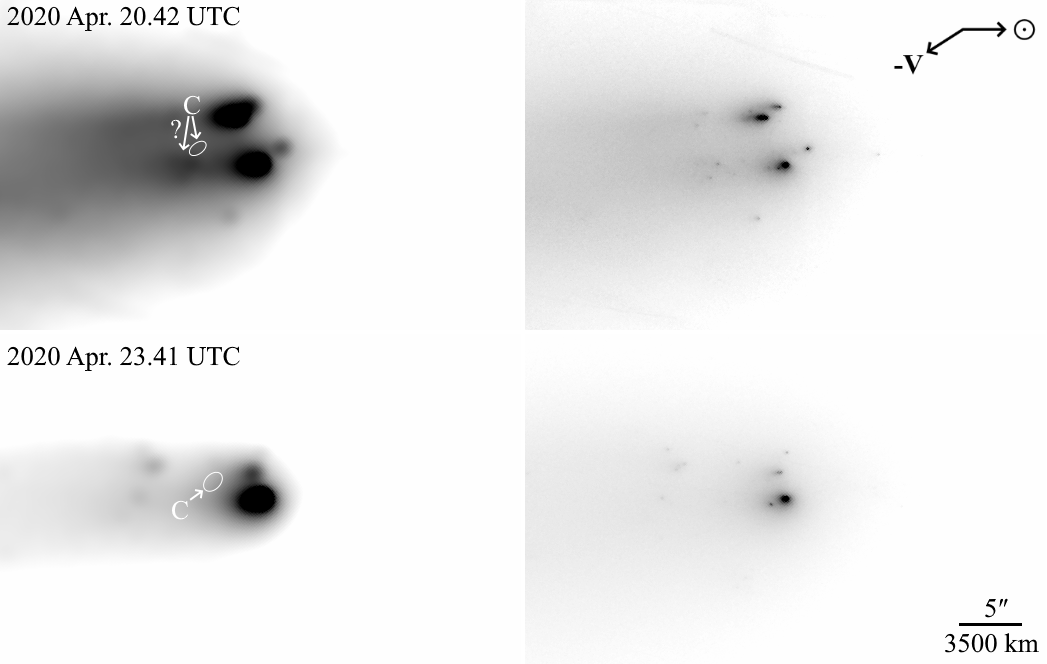}
\caption{The \hst~images convolved to a $2''$ FWHM similar to ground-based data (left column) and the original images (right column) showing the illusion of fragment C consists of several faint fragments.}
\label{fig:mosaic-conv}
\end{figure*}

Using $\sim$1,000 ground-based measurements submitted by observers world-wide, the Minor Planet Center (MPC) identified four fragments of Comet ATLAS, designated as C/2019 Y4-A through D\footnote{See Minor Planet Electronic Circular (MPEC) 2020-H28 (\url{https://minorplanetcenter.net/mpec/K20/K20H28.html}). An additional fragment E was assigned in MPEC 2020-J16 (\url{https://minorplanetcenter.net/mpec/K20/K20J16.html}), but was eventually linked to fragment B (MPEC 2020-K131, \url{https://minorplanetcenter.net/mpec/K20/K20KD1.html}).}. The MPC identification shows that these fragments became separated from the common parent around 2020 March 23 (fragment A), March 31 (B), April 6 (C), and April 9 (D), and were tracked until April 19 (A), May 10 (B), May 2 (C), and April 17 (D). Figure~\ref{fig:unc} shows the ephemeris positions of these fragments, as well as the correspondence for fragments A and B. The identification of fragments A and B on the images is straightforward and robust, as they are the brightest components in the system and are unambiguously close to the ephemeris nominal of the respective fragment. Fragments C and D, on the other hand, do not have clear correspondence on the images. Fragment D was last observed about 3 days before the \hst~observation, and its absence may simply indicate a complete disruption. The case of fragment C is more puzzling, as it was reportedly tracked from the ground until early May. Interestingly, if we convolve the \hst~images to $2''$ FWHM, comparable to typical ground seeing, a blob-like artifact appears near the ephemeris position of fragment C on the April 20 image (Figure~\ref{fig:mosaic-conv}). We also note that the true ephemeris error of the nearby bright fragment B ($\sim1''$--$2''$ as estimated from Figure~\ref{fig:unc}) is almost as large as the sky-plane distance between B and C. These observations support the idea that ground-based fragment C is a data artifact and highlight the importance of angular resolution on highly structured targets.

\begin{figure*}[!htb]
\centering
\includegraphics[width=\textwidth]{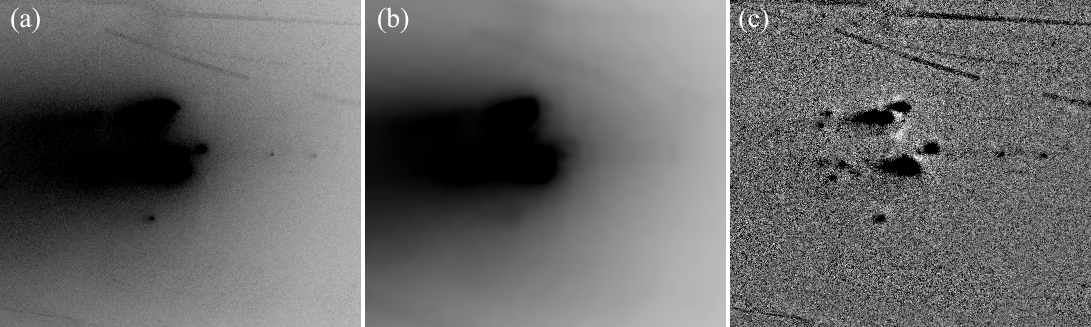}
\caption{Demonstration of coma removal process: (a) the original, median-combined composite image; (b) modeled coma generated by smoothing the image using a $15\times15$ pixels boxcar function; and (c) subtracted, coma-free image showing the fragments. All panels are scaled using the IRAF z-scale algorithm.}
\label{fig:sub}
\end{figure*}

The images show that fragments A and B consist of clusters of fragments moving at distinguishably different directions and speeds. Many of these fragments are embedded in the coma, and are too faint and/or too close to each other to be resolved from the ground, adding additional challenges to fragment identification and tracking. Hence, we take the following steps to identify and measure the fragments:

\begin{enumerate}
    \item We first blinked the exposures to identify groups of fragments that move with similar directions and speeds (i.e. the fragments would not trail after combining all frames from an orbit). Four groups are identified: clusters of fragments A and B, and two isolated fragments towards the tail-side. For each fragment group, we median combined the exposures in each set using the co-motion of the group, resulting in four composite images per orbit. Hence, each composite image is ``appropriated'' to one fragment group, as it is generated using the motion rate optimized for this group. Identification and measurement of fragments in a group was only done using the corresponding image of this group.
    
    \item We smoothed each composite image using a simple $15\times15$ pixels boxcar function then subtracted this from the original in order to suppress large-scale variations across the image (Figure~\ref{fig:sub}). We chose a FWHM of 15~pixels based on the typical apparent size of the fragments (estimated to be 5--10~pixels).
    
    \item We used {\tt Photutils} \citep{Bradley2020} to extract sources from the composite images. Sources were extracted using an aperture of 5-pixel radius and an empirical sigma level of $10\sigma$. To exclude image artifacts, we inspected the original frames and noted detections that correspond to prominent sources seen only  in one frame, and removed them from the detection list. In this way, we identified 23, 23 and 21 sources in the images from each orbit.
\end{enumerate}

We then derived the astrometry and photometry of each identified source. The absolute astrometry of each source was derived based on the recomputed astrometric solutions described in \S~\ref{sec:obs}. The photometry of each source was measured using a 1.5-pixel ($0.06''$) radius aperture and was then corrected for aperture losses, estimated using the standard point source function (PSF) model generated for WFC3/F350LP by {\tt TinyTim} \citep{Krist2011}. This approach minimizes the contamination from the coma and nearby sources. The local background estimated using a sigma-clipped median within an annulus with an inner and outer radius of 10 and 20~pixels ($0.4''--0.8''$). By looking at the scatter of the points from individual exposures, we estimated an uncertainty of $\sim1$~pixel ($0.04''$) for astrometry and $\sim0.1$~mag ($10\%$) for photometry.

\subsection{Fragment Tracking}
\label{sec:an:tracking}

\begin{figure*}[!htb]
\centering
\includegraphics[width=\textwidth]{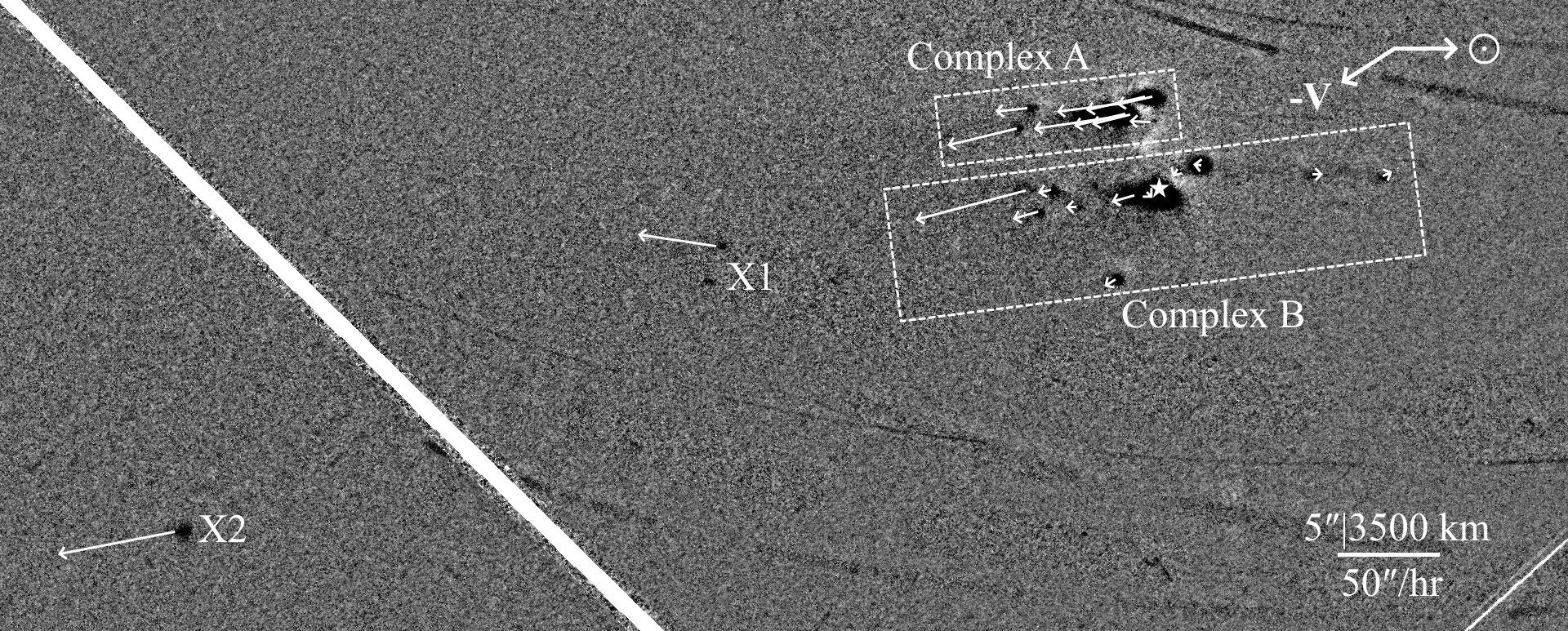}
\caption{Fragment tracks (vectors) on the first April 20 image. The vectors are enlarged by a factor of 10 for clarity. Note that there are some apparent sources on the image not marked by vectors (for example, a source just below X1): these are not detected following the procedure outlined in the main text and are likely image artifacts. The primary component used as a point of reference is marked by a star symbol. Also marked are the approximate boundaries of Complexes A and B (the fragment clusters corresponding to fragments A and B identified in ground-based data; see the discussion in the main text). X1 and X2 are two isolated fragments with no apparent association with any MPC-identified fragments.}
\label{fig:frag-vec}
\end{figure*}

Our next step was to link the sources between the orbits into unique tracklets. We first focused on the April 20 image pair and attempted to identify the same sources detected in the two orbits. Each orbit has 23 sources all successfully linked, as shown in Figure~\ref{fig:frag-vec}. Most (21 out of 23) of the fragments were located near MPC-identified fragments A and B. We hereafter refer to these two clusters as Complex A and Complex B. Two additional isolated fragments appear in the tailward direction of the main components, which we label as X1 and X2. These two fragments are not associated with any of the MPC-identified fragments, and are too faint ($V\sim23$) to be detected by most ground-based telescopes.

We then extend these short tracks to the image epoch on April 23 to match the sources detected on the latter date. The extrapolation is based on the sky-plane motion of the fragments assuming unaccelerated motion relative to the primary (the brightest component in the system). The change of viewing geometry from April 20 to 23 is minimal (the phase angle, or Sun--Comet--Earth angle, only increased by $2.7^\circ$) and is therefore neglected. The uncertainty range is derived assuming a measurement error of 1~pixel as previously estimated in \S~\ref{sec:an:id}, which translates to an extrapolation error of 48~pixels ($1.92''$) in radius on April 23. The Keplerian shear due to the ejection velocities is approximately $10^3$~km or $1.5''$, estimated using {\it vis-viva} equation taking the relative fragment speed of $\sim10$~m/s (as will be shown later in \S~\ref{sec:disc:1d}).

\begin{figure*}[!htb]
\centering
\includegraphics[width=\textwidth]{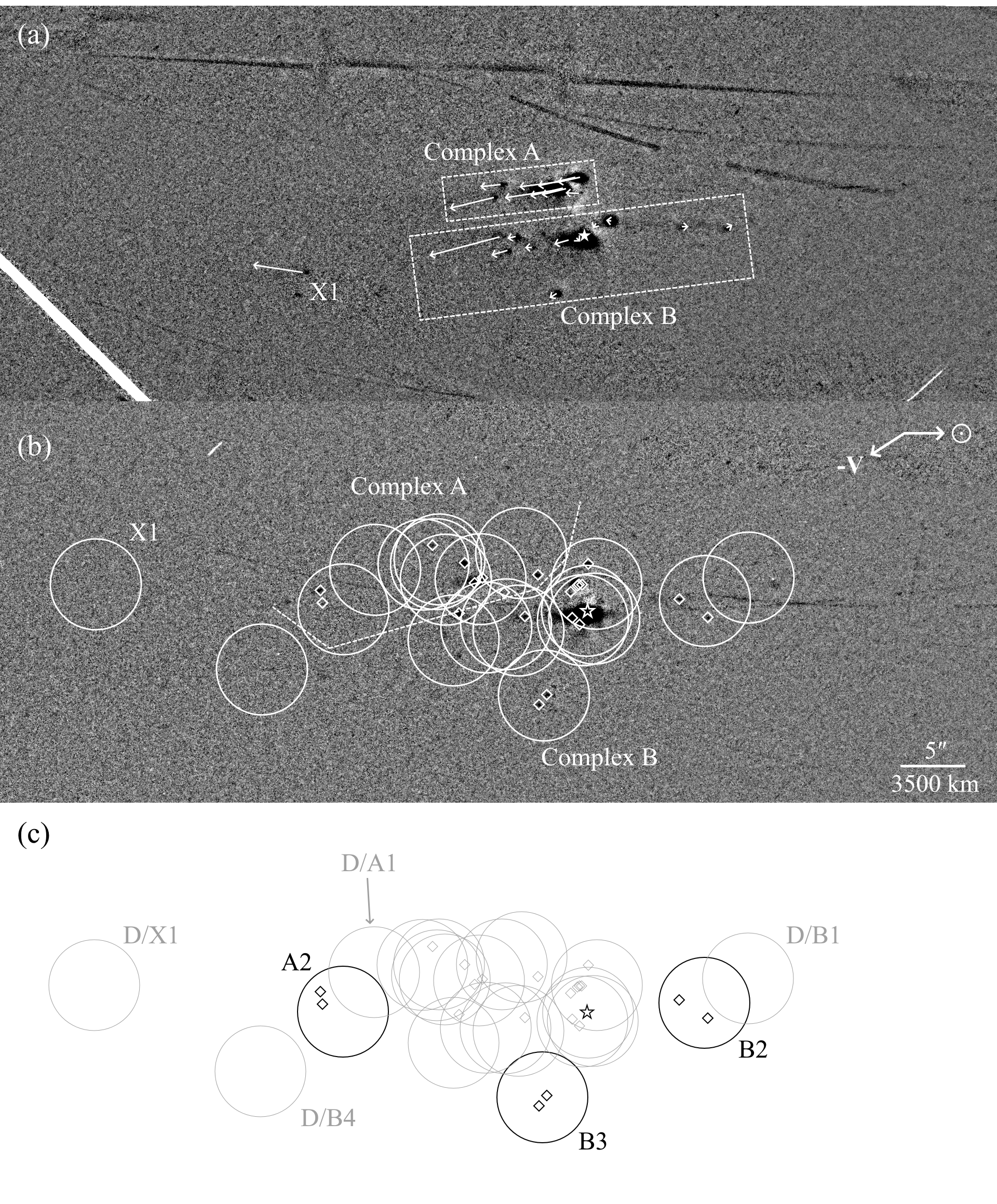}
\caption{A cropped version of Figure 5 (panel a), compared to the locations of the fragments on the 2020 April 23 image as predicted by a simple extrapolation based on their on-sky motions and Keplerian divergence (circles, with sizes representing the uncertainty ranges), as well as the actual detections on the image (diamonds), as in panels b--c. Fragment X2 is outside the image and is not shown. The primary component being used as the point of reference is marked by a star symbol. Panel b is overlaid with the coma-subtracted composite image; the dashed line in the upper panel marks the crude boundary between Complexes A and B. Panel c labels the identified track-able fragment clusters.}
\label{fig:frag-ext}
\end{figure*}

Figure~\ref{fig:frag-ext} shows the extrapolated locations of the fragments on the April 23 image as well as the actual detections. Besides the primary component which is used as the point of reference, none of the sources can be uniquely matched. The simplest explanation is that most of the fragments detected on April 20 have either split or completely disrupted in the intervening 3 days. An alternative scenario is that the fragments are pushed beyond the predicted circle by strong non-gravitational acceleration; such acceleration would need to be $a \sim 2s/t^2\approx1\times10^{-6}~\mathrm{au~d^{-2}}$, comparable to the value observed in disrupting comets \citep[e.g.][]{Hui2015}, again implying catastrophic disruption.

Although no unique identification can be made, we note that three pairs of fragments each can be grossly matched to a prediction circle, while four ex-fragments have apparently disappeared, as shown in the bottom panel of Figure~\ref{fig:frag-ext}. Each fragment pair or defunct fragment is labeled by a letter denoting the complex it belongs to plus a sequence number, such as A1. (Hereafter we use disappeared and defunct as synonyms.) To ease our discussion, a prefix of ``D/'' is added to the defunct fragments.

\section{Discussion}

\subsection{Fragment Size Distribution}
\label{sec:an:fsd}

The photometry provides a measure of the scattering cross-section, $C_e$, of each component.  The scattering cross-section is dominated by dust and hence provides only an upper limit to the sizes of the fragments. The radius of an equal-area circle, $r_e$, can be calculated by

\begin{equation}
    r_e = \left( \frac{C_e}{\pi} \right)^{1/2}
\end{equation}

\noindent where cross-section $C_e$ is given by 

\begin{equation}
\label{eq:c}
    C_e = \frac{\pi r_\mathrm{H}^2 \varDelta^2}{p_{\lambda} (1~\mathrm{au})^2 \phi(\alpha)} 10^{-0.4(m_{\lambda}-m_{\odot, \lambda})}.
\end{equation}

\noindent Here $p_{\lambda}=0.04$ is the typical geometric albedo of cometary nuclei \citep{Lamy2004}, $\phi(\alpha)$ is the Schleicher--Marcus (also called Halley--Marcus) phase function at phase angle $\alpha$ \citep{Schleicher1998, Marcus2007}, $m_{\lambda}$ is the measured magnitude for the specific band $\lambda$, and $m_{\odot, \lambda}=-26.8$ is the apparent brightness of the Sun in F350LP \citep{Willmer2018}\footnote{See also \url{http://mips.as.arizona.edu/~cnaw/sun.html}.}. Given a $\sim \pm0.1$~mag error in the photometry (\S~\ref{sec:an:id}), the derived $C_e$ are accurate to $\pm$10\%, although systematic errors (e.g.~due to the unknown dust albedo and color) may be substantially larger.

\begin{figure*}[!htb]
\centering
\includegraphics[width=\textwidth]{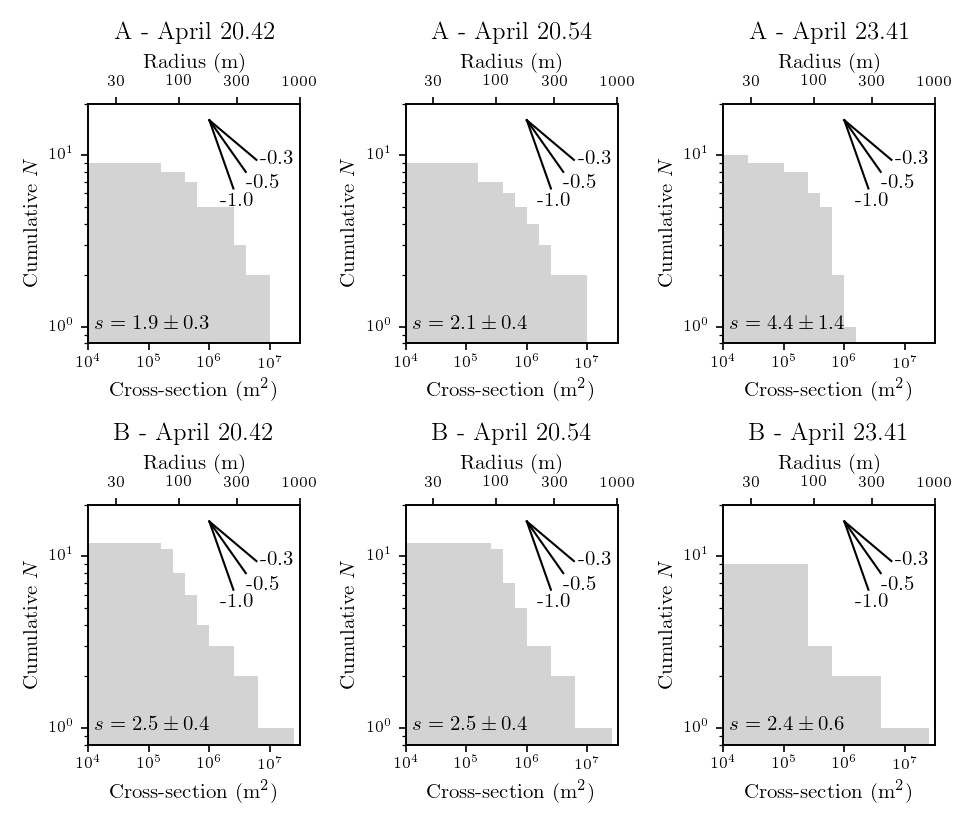}
\caption{Cumulative distributions of cross-section of the fragments at different epochs. Top 3 panels shows the distribution of Complex A, while bottom 3 panels shows that of Complex B. The gradients of power-law indices $1-g$ (assuming the differential distribution of the cross-section is written as $N(C_e) \propto C_e^{-g}$) are given in each panel for reference. The differential size index $s=2g-1$ for each panel derived by the \citet{Clauset2009} power-law fitting algorithm is also shown.}
\label{fig:frag-fsd}
\end{figure*}

The fragment cross-section distribution shown in Figure~\ref{fig:frag-fsd} reveals two interesting results:

\begin{itemize}
    \item At $r_e<$50--100~m or $V\gtrsim24$--25, the distribution is flat. The \hst/WFC3 Exposure Time Calculator\footnote{\url{http://etc.stsci.edu/etc/input/wfc3uvis/imaging}.} suggests that the images should reach a signal-to-noise ratio of 10 (a rather conservative threshold for source detection) for a $V=27.3$ ($r_e=20$~m) solar-spectrum object, seemingly indicating a paucity of small, faint fragments. However, the strong coma can make faint fragments invisible, effectively introducing a bias against smaller fragments. A quick examination shows that the coma region in the April 20 image has a limiting magnitude of $V\approx25$. The coma had faded considerably by April 23, coinciding with the increasing detections of fainter fragments, supporting the idea that the absence of small fragments is an observational bias.
    \item Complex A appears to be rapidly evolving, although the interpretation is somewhat undermined by the small number statistics. The differential index of the size distribution, $s$, can be calculated by $s=2g-1$, where $g$ is the differential index of the distribution of the cross-section that can be expressed as $N(C_e) \propto C_e^{-g}$. We derive $g$ by fitting the {\it cumulative} distribution of each Complex observed in each orbit using the algorithm described by \citet{Clauset2009} which provides us the cumulative index, $1-g$. Complex A has $s\approx2.0\pm0.4$ on April 20 and $s=4.4\pm1.4$ on April 23, showing a $1.3\sigma$ difference. The size distribution of Complex B remains largely constant ($s\approx2.5\pm0.5$) from April 20 to 23. The size distribution of Complex A on April 20 and Complex B is generally shallower than the numbers measured on other fragmenting comets, such as 73P/Schwassmann-–Wachmann 3 \citep[$s=3.34\pm0.05$,][]{Ishiguro2009}, 332P/Ikeya--Murakami \citep[$s=3.6\pm0.6$,][]{Jewitt2016}, and the deca-meter fragments in the Kreutz family \citep[$s=3.2$,][]{Knight2010}, while that of Complex A on April 23 appears steeper (but not statistically significant). In this regard, we note that fragment kinematics (to be discussed in \S~\ref{sec:disc:1d}) also shows that Complex A appears to be much shorter-lived compared to Complex B, despite the two having a similar integrated brightness at the beginning (see \S~\ref{sec:disc:3d}). This suggests that Complex A is the product of a small, short-lived, and much more active fragment compared to Complex B. The $r_e$ of the largest piece in Complex A is $\sim500$~m. Considering all the uncertainties, the entire pre-fragmentation nucleus of Comet ATLAS is likely no much larger than a few $10^2$~m in radius.
\end{itemize}

\subsection{Size and Dynamical Evolution of the Fragments}

\subsubsection{Evolution Over 3 Hours On 2020 April 20}
\label{sec:disc:1d}

\begin{figure*}[!htb]
\centering
\includegraphics[width=\textwidth]{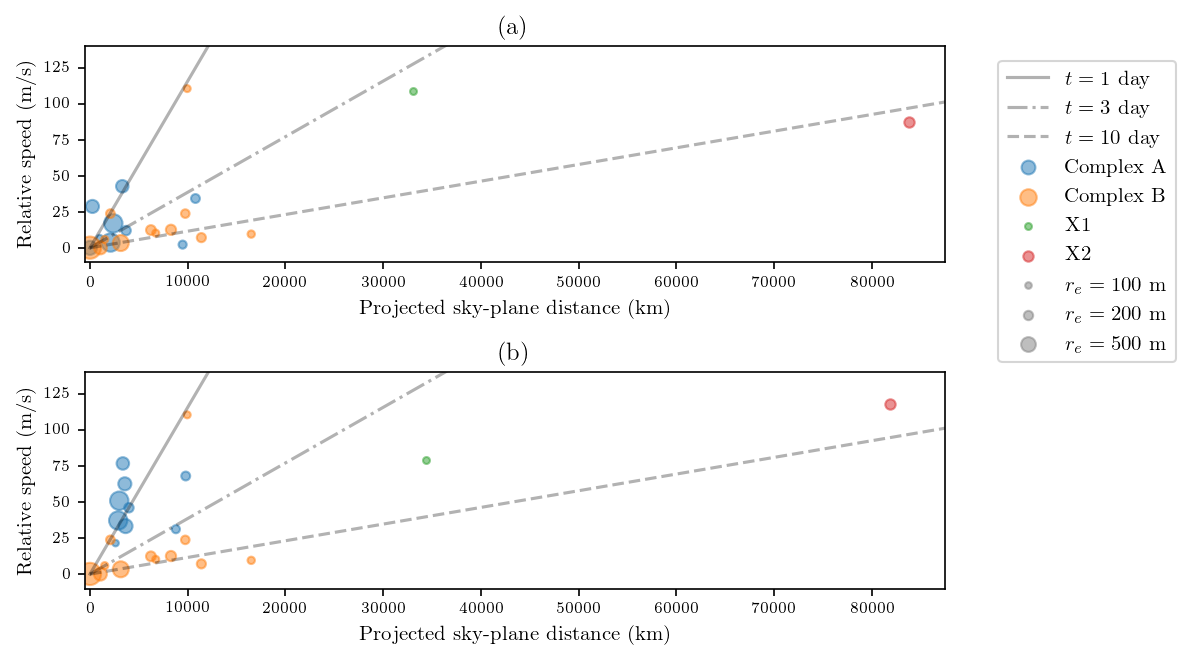}
\caption{Fragment sizes, sky-plane speeds and the projected distances of the fragments measured on the 2020 April 20 image pair. The speeds and distances are measured with respect to different points of origin: panel (a) assumes multiple origins of the fragments: the ones in Complex A with respect to the primary in Complex A, the ones in Complex B with respect to the primary in Complex B, and X1, X2 with respect to Complex A; while panel (b) assumes a single origin of all fragments, the primary component in Complex B. The primaries are defined as the brightest component in the respective Complex. Uncertainties are too small to show on this plot. Fragment sizes are inferred from the mean brightness (with the change in brightness neglected). Solid, dashed and dashed-dotted lines are the lapsed times since ejection assuming unaccelerated fragments.}
\label{fig:frag-dist}
\end{figure*}

Assuming that the fragments do not accelerate, the time-of-flight, $t$, can be derived from $t=l/v$, where $l$ is the projected distance between the fragments and $v$ is the relative sky-plane speed. Fragments ejected simultaneously will form a straight line on the speed--distance plot. Figure~\ref{fig:frag-dist} shows the relation between fragment sizes, relative sky-plane speeds and distances between the fragments as measured with the 2020 April 20 image pair. The impact caused by the change in observing geometry is negligible over 3~hours.

The large scatter in the data suggests that the fragments were released over a range of times, arguing against an impulsive origin of the fragments. Regardless of where the fragments originated, Complex A appears to be very young, with most members $\lesssim3$~days in age, suggesting that this part of the system was evolving rapidly. Complex B and possibly fragment X2 appears to be older, with an age of $\sim2$~weeks. This age is in line with the first reports of the fragmentation which were made 2~weeks before the \hst~observation \citep{Steele2020, Ye2020b}, suggesting that these are likely the first-generation fragments produced during the initial fragmentation.

The origin-dependent plots provide insights into the fragmentation history of the system: Figure~\ref{fig:frag-dist}a assumes all fragments in A, X1 and X2 were ejected from A, and those in B were ejected from B, whereas Figure~\ref{fig:frag-dist}b assumes all fragments were ejected from B. Scenario (b) reveals an extremely high ejection speed of $\sim30$--80~m/s and a very young age of Complex A (less than a day). An ejection speed of tens of~m/s is very high compared to separation speeds of most fragmenting comets \citep[only a few~m/s or less, cf.][]{Sekanina1982}. Although scenario (a) also implies high ejection speed, the speeds in scenario (b) are $\sim5$ times higher and require $5^2=25$ times more energy to attain, making the scenario less likely. Additionally, ground observations made a few days preceding the \hst~observations revealed fragments resembling Complex A, hence the age of Complex A could not have been younger than a few days. All these evidence favor scenario (a) which implies that the members in Complex A were tertiary products of the fragmentation.

However, even with scenario (a), the relative speeds between fragments are still larger than the typical value observed in many other fragmenting comets, which is puzzling. Scenarios that could produce fast fragments include self-propelled acceleration and explosion on the nucleus \citep{Stevenson2010}. Figure~\ref{fig:frag-vec} shows that most fragments travelled in the Sun--comet direction which is consistent with sublimation-driven acceleration. The force produced by non-central sublimation can be written as

\begin{equation}
    F \sim k_\mathrm{R} f_s \pi r_\mathrm{N}^2 v_\mathrm{th}
\end{equation}

\noindent where $k_\mathrm{R} \approx 0.5$ is the dimensionless momentum transfer coefficient \citep[cf.][]{Crifo1987, Attree2019, Jewitt2020}, $f_s$ is the specific sublimation rate at the surface, $r_\mathrm{N}$ is the effective radius of the fragment, and $v_\mathrm{th} \sim 10^3~\mathrm{m/s}$ is the typical gas outflow speed.  The value of $f_s$ was calculated using the model described by \citet{Cowan1979}\footnote{A web-based tool can be found at \url{https://pds-smallbodies.astro.umd.edu/tools/ma-evap/index.shtml}.}, in which we obtain $f_s\sim1\times10^{-4}~\mathrm{kg~m^{-2}~s^{-1}}$ assuming H$_2$O as the dominating species. Given that acceleration $a=F/m$ where $m$ is the fragment mass, we have

\begin{equation}
    a = \frac{3 k_\mathrm{R} f_s v_\mathrm{th}}{4 \rho r_\mathrm{N}}
\end{equation}

\noindent where $\rho=500~\mathrm{kg~m^{-3}}$ is the assumed bulk density of the comet. Since the fragments travelled $\sim10^4$~km in $\sim2$~weeks (estimated from Figure~\ref{fig:frag-dist}), the acceleration would be $a=2s/t^2=10^{-5}~\mathrm{m~s^{-2}}$, implying a maximum initial fragment diameter of $r_\mathrm{N}=8$~m. (We note that this does not contradict the detection limit of $r_e\sim50$~m established in \S~\ref{sec:an:fsd}, as $r_e$ is derived from the dust-contaminated cross-section and is only an upper limit to the sizes of the fragments.) This derived acceleration is also grossly consistent with the value estimated in \S~\ref{sec:an:tracking}. Such fragment size is possible given the contamination of dust within the photometry aperture. Hence, self-propulsion of fragments by asymmetric outgassing forces could explain the observed high speeds of the fragments. We will come back to this issue again in \S~\ref{sec:disc:mech}.

\begin{figure*}[!htb]
\centering
\includegraphics[width=\textwidth]{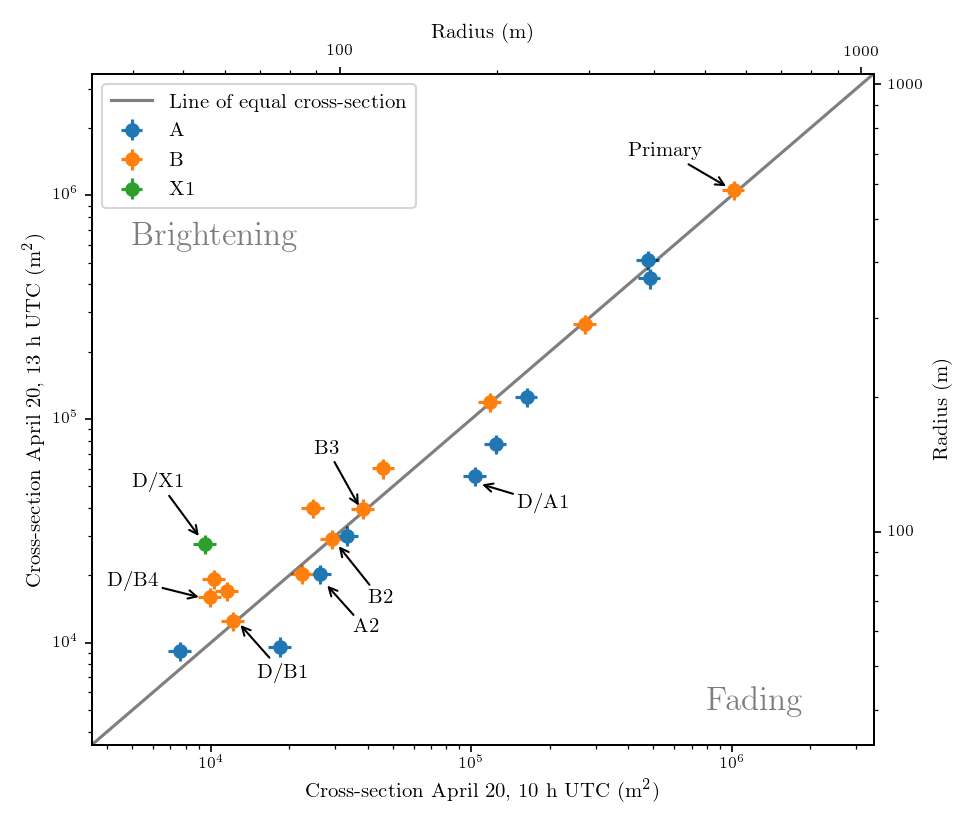}
\caption{Changes of cross-section areas (a proxy of brightness) of Complexes A, B and X1 between the two orbits on April 20. The diagonal line is the line of equal cross-section area that separates the ``brightening'' regime (upper-left corner) and ``fading'' regime (lower-right corner). We estimate a 10\% uncertainty of the data points following the discussion in \S~\ref{sec:an:id}. Also labeled are the fragments trackable to April 23 (cf. \S~\ref{sec:an:tracking}).}
\label{fig:frag-crosssection-evolv-apr20}
\end{figure*}

Figure~\ref{fig:frag-crosssection-evolv-apr20} shows the changes of cross-section  of the fragments between the two orbits on April 20. The impact of the change in observing geometry ($\sim1\%$) is smaller than the uncertainty and is therefore negligible. The seven fragments ``traceable'' to April 23 (including the disappeared ones) show an interesting tendency: the three fragments survived to April 23 tend to have near-constant brightness, while the ones that disappear tend to have fluctuating brightness ($\gg 10\%~\mathrm{hour^{-1}}$) except for D/B1. Further interpretation is limited by the small statistics, but this is in line with the general understanding of a disrupting cometary nucleus, that its brightness can fluctuate greatly due to the rapid release and dissipation of dust and gas.

\subsubsection{3-Day Evolution From 2020 April 20 To 23}
\label{sec:disc:3d}

\begin{figure*}[!htb]
\centering
\includegraphics[width=\textwidth]{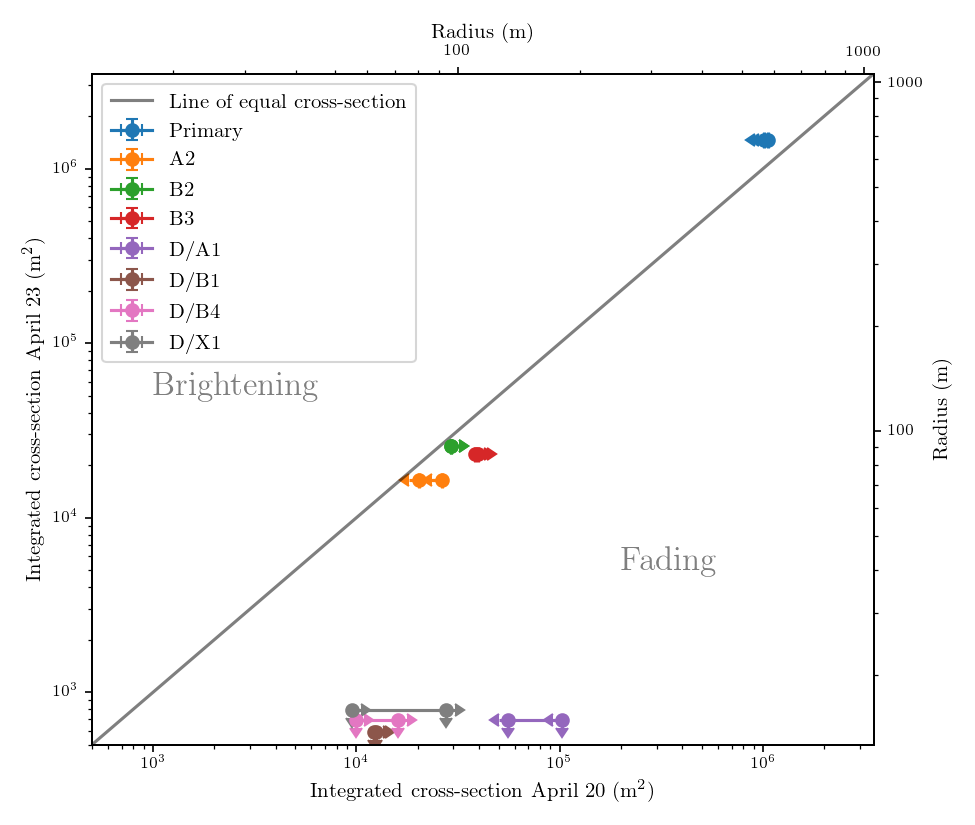}
\caption{Changes of cross-section areas (a proxy of brightness) of Complexes A, B and X1 from 2020 April 20 to 23. Arrows along the X-axis indicate the direction of the change measured on the April 20 data (brightening or fading). The diagonal line is the line of equal cross-section that separates the ``brightening'' regime (upper-left corner) and ``fading'' regime (lower-right corner). The 10\% uncertainty of the data points (see \S~\ref{sec:an:id}) is too small to be shown. The detection limit of the April 23 image is $\sim700~\mathrm{m^2}$, but the non-detections (the data points with Y values below $10^3~\mathrm{m^2}$) are offset slightly for clarity.}
\label{fig:frag-crosssection-evolv}
\end{figure*}

Figure~\ref{fig:frag-crosssection-evolv} shows the change of cross-section areas (brightness) of the seven traceable fragments from April 20 to 23. In addition to the fact that the most strongly fluctuating fragments have all disappeared, the plot also shows a lack of correlation between fragment survivability and the initial brightness: for example, D/A1 is the second-brightest fragment in the list but has disappeared, while some fainter fragments have survived. However, as noted above, fragment brightness is not a direct measure of the physical size of the fragment due to the contamination of near-nucleus dust. The actual size of A1 could be much smaller than B3, but only appear as bright since it was more actively releasing dust.

Another interesting phenomenon is that the surviving fragments all appear to have split into doubles, as shown in Figure~\ref{fig:frag-ext}. Split comets will gain more cross-section area, as the cross-section area decreases more slowly than does the volume. If we split a spherical comet with a radius of $r$ into $N$ equal spherical pieces, the total cross-section area obeys $\propto N^{1/3}$. Hence comet split will increase the total cross-section areas by 26\% assuming spherical fragments and neglecting dust contamination. However, Figure~\ref{fig:frag-crosssection-evolv} shows that the total brightness of these fragments has instead shrunk by up to 40\%, implying either a significant loss of material (30--70\% in mass) or a reduced activity after the split.

Assuming the classic $r_\mathrm{H}^{-4}$ law (as implied by Equation~\ref{eq:c}), Comet ATLAS as a whole should have brightened by 0.3~mag from April 20 to 23, but in reality it has faded by about 0.2~mag, corresponding to a secular fading rate of $\sim0.2~\mathrm{mag/day}$. The fading is primarily caused by the fading of Complex A, which has faded by 40\%. The brightness of Complex B is largely constant, though this is largely due to the brightening of the primary component making up for the fading of other fragments (which collectively faded by 50\%). This implies that all these fragments except for the primary fragment would disappear in a few days, consistent with subsequent ground observations.

\subsection{Fragmentation Mechanism}
\label{sec:disc:mech}

Cometary fragmentation can be caused by a number of processes, such as tidal stress, rotational instability, explosion due to internal gas pressure, and external impacts \citep[cf.][]{Boehnhardt2004}. For the case of Comet ATLAS, the orbital configuration argues against tidal stress as a possible cause, since the comet has $q=0.25$~au (well outside the Roche limit of the Sun) and an inclination $i=45^\circ$ (which keeps it away from major planets). The dispersion in fragment time-of-flight (as shown in Figure~\ref{fig:frag-dist}) also argues against an impulsive origin of the fragmentation. The two remaining possibilities are rotational instability and explosion through intense gas pressure.

Rotational instability arises from asymmetric outgassing of the comets, which can cause them to spin up and exceed the centripetal limit. Following the discussion in \citet[][\S~3]{Jewitt2016}, the characteristic timescale for a cometary nucleus to become rotationally excited is

\begin{equation}
\label{eq:spinup1}
    \tau_s = \frac{\rho r_\mathrm{N}^2}{v_\mathrm{th} k_\mathrm{T} f_A f_s P}
\end{equation}

\noindent where $\rho$, $v_\mathrm{th}$, $r_\mathrm{N}$ follow the definitions given above, $k_\mathrm{T}\approx0.005$ is the dimensionless moment arm \citep{Belton2011}, $f_A$ is active fraction of the nucleus, $f_s$ is the specific sublimation rate at the surface, and $P$ is the rotation period of the nucleus, all in International System of Units where applicable. The active fraction $f_A$ varies greatly among comets, with a general lower limit of $\sim1\%$ \citep{Ye2016}, but can approach $100\%$ for very active comets (such as the ones in fragmentation). The value of $f_s$ is solved using the model described by \citet{Cowan1979}, from which we obtain $f_s\sim1\times10^{-4}~\mathrm{kg~m^{-2}~s^{-1}}$ depending on the mode of the sublimation (subsolar or isothermal) assuming H$_2$O as the dominating species. Assuming $P=6$~hr \citep[cf.][]{Samarasinha2004}, Equation~\ref{eq:spinup1} can be further simplified to

\begin{equation}
\label{eq:spinup2}
    \tau_s \approx (0.0001\mathrm{-}0.05) r_\mathrm{N}^2~\mathrm{[days]}
\end{equation}

For fragments at $r_\mathrm{N}=50$~m, $\tau_s=0.3$--100~days. In particular, a larger $f_A$ (as can be expected for a fragmenting nucleus) would result in a smaller $\tau_s$ of around the order of a day, in line with the observed lifetime of small fragments. For the primary fragment ($r_\mathrm{N}=1000$~m), $\tau_s$ ranges from a few months to several decades. Interestingly, our unsuccessful post-perihelion recovery attempts (to be reported in a separate paper) indicate that the primary component only survived for another 1--2 months, also consistent with the derived $\tau_s$.

Rotational disruption appears to be able to explain the destruction of the observed high-order fragments, but can it be responsible for the initial fragmentation? The size of the original nucleus of Comet ATLAS is uncertain, but is likely no larger than a few kilometers, given that the primary component has $r_\mathrm{N}<1000$~m (\S~\ref{sec:an:fsd}). Accounting for a lower $f_A$ as well as $f_s$ at a larger distance from the Sun, we derive $\tau_s\gtrsim1$~yr for a km-class body, which is not very constraining. On the other hand, the surface escape velocity of a km-class body is around the order of 1~m/s. Our calculation in \S~\ref{sec:disc:1d} has showed that the fragments could be ejected slowly and then self-propelled to high speeds. Therefore, it is possible that the initial fragmentation was also driven by rotational disruption.

Another possible scenario is an energetic blow-off caused by high internal gas pressure. It has been hypothesized that runaway sublimation of sub-surface supervolatiles can create high gas pressure under the surface, which can lead to explosion if the gas cannot be released through surface activity \citep[e.g.][]{Kuehrt1994}. This hypothesis is also consistent with the significant bluing of the comet observed shortly before its fragmentation as observed by \citet{Hui2020} which indicated the release of a large amount of gas, though we note that rotational disruption can also expose the sub-surface ices which can lead to the same phenomenon.

The thermal skin depth $s_\mathrm{th}$ can be estimated by

\begin{equation}
    s_\mathrm{th} \sim (\kappa \tau) ^ {1/2}
\end{equation}

\noindent where $\kappa=3\times10^{-7}~\mathrm{m^2~s^{-1}}$ is the thermal diffusivity, and $\tau\sim1$~yr is the time that the comet has been heated. Inserting these numbers, we derive $s_\mathrm{th}=3$~m.  Given this, the question arises as to how the comet could survive at least one perihelion passage at $q=0.25$~au with ice so close to the surface. 

We postulate that Comet ATLAS is a chunk from the interior of its progenitor, and that it separated post-perihelion. The ices would be shielded by the outer crust of the progenitor during the past perihelion, and would have remained unperturbed. Given that the fragmentation of Comet ATLAS occurred at $r_\mathrm{H}=1.5$~au, the fragmentation of the progenitor would need to take place at $r_\mathrm{H}\gg1.5$~au post-perihelion in order to satisfy the assumption that the ices are unperturbed by the solar heat. We note that this does not contradict the broad constraint set by \citet{Hui2020} which suggested that the fragmentation of the progenitor likely occurred within $r_\mathrm{H}\lesssim10$~au. Fragmentation at large heliocentric distance might not be an unusual occurrence: \citet{Sekanina2007} suggested that many Kreutz comets likely separated from each other well before/after the perihelion passage.

Also, fragments produced from the blown-off outer crust of the progenitor would contain less supervolatile ices and have a higher chance to survive the next perihelion. To that end, C/1844 Y1, Comet ATLAS' sibling, has survived the perihelion without any apparent fragmentation \citep{Kronk2003}. Could C/1844 Y1 be the outer crust of the progenitor? A ``drier'' nucleus can increase the chance of the nucleus surviving a small perihelion passage, but so does a large nucleus. The size of C/1844 Y1 has not been previously reported, but can be crudely constrained using the relation between comet brightness and H$_2$O production rate derived by \citet{Jorda2008}. Taking a near-peak brightness of $V\sim2$~mag (reported by J. Robinson on 1844 December 23) and active surface fraction of $>10$\% (appropriated for a near-Sun comet), we derive a nucleus diameter of $<6$~km, grossly comparable to the upper limit of the size of Comet ATLAS that we estimated earlier. However, given that the size constraints of both comets are only upper limits, we cannot rule out the possibility that C/1844 Y1 being much larger than Comet ATLAS, therefore no definite conclusion on the nature of C/1844 Y1 can be made.

Nevertheless, fragments derived from a non-uniform progenitor will likely exhibit different observational properties and behavior. Drier fragments may have a smaller active fraction and a smaller observable turn-on distance due to the paucity of near-surface volatiles. (C/1844 Y1 was discovered post-perihelion, therefore its turn-on distance is unknown.) They are also likely to be more resistant to spin-up excitation and other disruption mechanisms since that the sublimation activity is lower. We note that the Kreutz comets have exhibited behavior consistent with a non-uniform makeup: C/2012 E2 (SWAN) has a turn-on distance much larger than others, possibly hinting a nucleus that is substantially more volatile-rich than others \citep{Ye2014}. This could be analogous to the case of Comet ATLAS and C/1844 Y1 if the latter is indeed drier, though we note that C/SWAN and most Kreutz comets are likely about an order of magnitude smaller than Comet ATLAS and C/1844 Y1. Future discovery and characterization of members in this group will provide more insight to the makeup and fragmentation history of the progenitor.

\subsection{Comparison With C/1999 S4 (LINEAR)}

Besides Comet ATLAS, the only other disintegrated LPC studied by \hst~is C/1999 S4 (LINEAR)\footnote{\hst~also observed the fragmentation of C/1996 B2 (Hyakutake) and ill-fated C/2012 S1 (ISON), but Comet Hyakutake did not completely disintegrated, and Comet ISON's demise was after the \hst~observation.}. Table~\ref{tbl:comp} summarizes the measured and derived properties of the two comets.

\begin{table*}
\centering
\caption{Measured and derived properties of C/1999 S4 (LINEAR) and C/2019 Y4 (ATLAS).}
\footnotesize
\begin{tabular}{ccc}
\hline
Property & C/1999 S4 (LINEAR) & C/2019 Y4 (ATLAS) \\
\hline
Orbit class & Dynamically new\tablenotemark{a} & Dynamically old\tablenotemark{b} \\
$q$ & 0.77~au\tablenotemark{a} & 0.25~au\tablenotemark{b} \\
Pre-disintegration diameter & 0.2--0.5~km\tablenotemark{c,d} & A few 0.1~km \\
First major disruption & $\sim20$~d pre-perihelion, at $r_\mathrm{H}=0.9$~au & $\sim60$~d pre-perihelion, at $r_\mathrm{H}=1.4$~au \\
Complete disintegration & $r_\mathrm{H}=0.7$~au (near perihelion) & $r_\mathrm{H}=0.5$~au (before perihelion) \\
$s$ & 3.7\tablenotemark{d} & Variable, from 2.0 to 4.4 \\
Polarization mode & High maximum\tablenotemark{e} & High maximum\tablenotemark{g} \\
Disrupted due to tidal stress? & Unlikely\tablenotemark{h} & Unlikely\tablenotemark{i} \\
Disrupted due to rotational instability & ? & Maybe\tablenotemark{i} \\
Disrupted due to gas pressure? & Maybe\tablenotemark{j} & Maybe\tablenotemark{i} \\
Disrupted due to external impact? & Unlikely\tablenotemark{h} & Unlikely\tablenotemark{h} \\
Note & Low CO abundance\tablenotemark{k,l} & Bluing shortly before fragmentation\tablenotemark{m} \\
\hline
\end{tabular}
\tablenotetext{a}{JPL solution \#99.}
\tablenotetext{b}{JPL solution \#13.}
\tablenotetext{c}{\citet{Weaver2001}.}
\tablenotetext{d}{\citet{Makinen2001}.}
\tablenotetext{e}{\citet{Kidger2002}.}
\tablenotetext{f}{\citet{Hadamcik2003}.}
\tablenotetext{g}{\citet{Zubko2020}.}
\tablenotetext{h}{These possibilities are not formally discussed in any published works, but we believe that they are unlikely given the high inclination of the two comets.}
\tablenotetext{i}{This work.}
\tablenotetext{j}{\citet{Samarasinha2001}.}
\tablenotetext{k}{\citet{BockeleMovan2001}.}
\tablenotetext{l}{\citet{Mumma2001}.}
\tablenotetext{m}{\citet{Hui2020}.}
\label{tbl:comp}
\end{table*}

Compared to Comet ATLAS, Comet LINEAR is a dynamically new comet likely on its first visit to the inner Solar System \citep{Farnham2001}. It also has a somewhat larger $q$ of 0.77~au. The start of the disintegration of Comet LINEAR was fortuitously captured by an \hst~target-of-opportunity program \citep[General Observer program 8276,][]{Weaver2001} on 2000 July 4, when the comet was at $r_\mathrm{H}=0.86$~au, considerably closer than the first major disruption of Comet ATLAS ($r_\mathrm{H}\sim1.4$~au). The differences in the start of the disruption, together with the dynamical properties of the two comets, appears to imply that the bulk strength of Comet ATLAS is likely to be significantly weaker than Comet LINEAR. This is interesting considering that Comet ATLAS, as a dynamically old and a returning comet that has passed close to the Sun at least once, should tend to be stronger in strength than a typical dynamically new comet such as Comet LINEAR. The two comets otherwise share a number of similar characteristics and behavior: both are likely sub-kilometer in size \citep{Makinen2001, Weaver2001}, become completely disrupted around 0.5--0.7~au, and exhibit a high-maximum polarization mode \citep{Hadamcik2003, Zubko2020}.

The break-up mechanism and composition of these two comets is also worthy of note. \citet{Samarasinha2001} suggested that the disintegration of Comet LINEAR was caused by internal explosion due to the sublimation of sub-surface supervolatiles such as CO. They showed that this was theoretically possible even with the extremely low CO abundance reported by ultraviolet and mid-infrared observations \citep{Mumma2001, Weaver2001}. However, it would appear unusual for this mechanism to operate on Comet LINEAR while comets with much higher CO abundances do not disrupt like this. Additionally, \citet{Mumma2001} showed that Comet LINEAR is deficient in other supervolatile species. As for Comet ATLAS, no direct CO measurement has been reported as of this writing. The significant bluing observed by \citet{Hui2020} is in line with a rapid increase of gas production, but broadband photometry does not permit distinguishing the contribution of different species, hence no strong conclusion can be made.

\section{Summary}

We presented high-resolution \hst~observations of disintegration of dynamically-old Near-Sun Comet C/2019 Y4 (ATLAS) between 2020 April 20 and 23. The data fortuitously covered a critical event: the demise of one of the two main fragment complexes. Our findings are as follows:

\begin{enumerate}
    \item Our observations showed that fragments C/2019 Y4-A and C/2019 Y4-B, originally identified as two single fragments in ground-based data, were actually two fragment clusters. Each fragment cluster consists of a few dozen fragments down to $\sim10$~m size.
    \item Complex A was clearly in the process of a complete disintegration, while Complex B stayed largely intact, despite the fact that they had a similar initial brightness. Photometry showed that Complex A lost $\sim70\%$ mass from April 20 to 23. The size distribution of Complex A also changed by $1.3\sigma$ during this period, indicative of rapid evolution of the system, while that of Complex B stayed largely the same. Fragment kinematics suggested that Complex A was only a few days old while Complex B was $\sim2$~weeks old. We speculate that Complex A was produced by a small (likely 100-m-class) and rapidly disrupting fragment.
    \item The fragments moved at on-sky speeds of $\sim10$~m/s, much higher than typical fragmenting comets. This could be explained either by self-propelled fragments ejected by centripetal disruption or an explosive blow-off due to internal gas pressure. The destruction of smaller fragments occurred within a timescale of a few days, consistent with spin-up disruption of small cometary nuclei at such sizes.
    \item The significant bluing of the comet observed shortly before the main fragmentation event \citep{Hui2020} indicates a release of a large amount of gas, which hints at a fast and large-scale sublimation of volatile ices.
    \item Comparison with disintegrated dynamically-new comet C/1999 S4 (LINEAR) shows that the bulk strength of Comet ATLAS is likely to be significantly weaker than Comet LINEAR, contrary to population statistics.
    \item We speculate that Comet ATLAS was derived from the ice-rich interior of a non-uniform progenitor that broke apart $\sim5$~kyr ago. The break-up may have occurred at $r_\mathrm{H}\gg1.5$~au in the outbound leg in order to shield the interior ice from the intense solar heat at $q=0.25$~au. The other major fragment produced by the break-up, C/1844 Y1, appeared to have survived its perihelion passage and could represent the ``drier'' part of the progenitor, though this explanation is undermined by the limited amount of data.
\end{enumerate}

\acknowledgments

We thank two anonymous referees for their review, as well as Matthew Knight for helpful discussion and comments. This research is based on observations made with the NASA/ESA {\it Hubble Space Telescope} obtained from the Space Telescope Science Institute, which is operated by the Association of Universities for Research in Astronomy, Inc., under NASA contract NAS 5-26555. Support for this work was provided by NASA through grant numbers HST-GO-16089 and 16111 from the Space Telescope Science Institute. JA and YK acknowledge funding by the Volkswagen Foundation. JA's contribution was made in the context of ERC Starting Grant 757390 CAstRA. 

This research made use of {\tt Montage} and {\tt Photutils}. {\tt Montage} is funded by the National Science Foundation under Grant Number ACI-1440620, and was previously funded by the National Aeronautics and Space Administration's Earth Science Technology Office, Computation Technologies Project, under Cooperative Agreement Number NCC5-626 between NASA and the California Institute of Technology. {\tt Photutils} is an {\tt Astropy} affiliate package for detection and photometry of astronomical sources.


%

\vspace{5mm}
\facilities{HST(WFC3)}


\software{{\tt L.A. Cosmic} \citep{vanDokkum2001}, {\tt Montage} \citep{Jacob2010}, {\tt Photutils} \citep{Bradley2020}, {\tt SCAMP} \citep{Bertin2010}, {\tt TinyTim} \citep{Krist2011}}




\bibliography{sample63}{}
\bibliographystyle{aasjournal}



\end{CJK*}
\end{document}